\documentclass[3p,times]{elsarticle}

\usepackage{ecrc}

\volume{00}

\firstpage{1}

\journalname{Physics Letters B}

\runauth{Kalita \& Sarmah}

\jid{plb}

\jnltitlelogo{Physics Letters B}

\CopyrightLine{2022}{Published by Elsevier Ltd.}

\usepackage{latexsym}
\usepackage{amssymb, amsmath, bm, physics}
\usepackage{subcaption}
\usepackage{mathptmx}
\usepackage{flushend}
\usepackage[colorlinks,citecolor=blue,urlcolor=blue,linkcolor=blue]{hyperref}
\usepackage{aas_macros}
\usepackage{orcidlink}

\biboptions{sort&compress}

\begin{document}

\begin{frontmatter}



\dochead{}

\title{Weak-field limit of $f(R)$ gravity to unify peculiar white dwarfs}


\author{Surajit Kalita\orcidlink{0000-0002-3818-6037}}
\address{Department of Physics, Indian Institute of Science, Bangalore 560012, India}
\ead{surajitk@iisc.ac.in}

\author{Lupamudra Sarmah\orcidlink{0000-0003-1651-9563}}
\address{Department of Physics, Indian Institute of Technology (BHU), Varanasi 221005, India}
\ead{lupamudrasarmah.phy20@itbhu.ac.in}

\begin{abstract}
In recent years, the idea of sub- and super-Chandrasekhar limiting mass white dwarfs (WDs), which are potential candidates to produce under- and over-luminous type Ia supernovae, respectively, has been a key interest in the scientific community. Although researchers have proposed different models to explain these peculiar objects, modified theories of Einstein's gravity, particularly $f(R)$ gravity with $R$ being the scalar curvature, seems to be one of the finest choices to explain both the regimes of these peculiar WDs. It was already shown that considering higher-order corrections to the Starobinsky model with two parameters, the structure of sub- and super-Chandrasekhar progenitor WDs can be explained self consistently. It is also well-known that WDs can be considered Newtonian objects because of their large size. In this paper, we derive the weak-field limit of $f(R)$ gravity, which turns out to be the higher-order correction to the Poisson equation. Later, we use this equation to obtain the structures of sub- and super-Chandrasekhar limiting mass WDs at various central densities incorporating just one model parameter.
\end{abstract}

\begin{keyword}
modified gravity \sep white dwarfs \sep Chandrasekhar limit
\PACS 04.50.Kd \sep 97.20.Rp \sep 97.10.Nf \sep 04.40.Dg

\end{keyword}

\end{frontmatter}


\section{Introduction}
Over the years, several theoretical, as well as phenomenological outcomes, have served as motivations to consider alternatives to general relativity (GR). As far as weak gravitational background is concerned, GR can quite accurately explain several phenomena, starting from the perihelion precession of Mercury to the prediction of black holes and gravitational waves~\citep{2004sgig.book.....C}. However, the disagreements between observations and expected results, leading to deviations from GR at the high-density regime with large scalar curvature $R$, demand significant modifications to this theory. Recent cosmological observations suggest that even though GR is well tested at the solar system level, it is not sufficient to explain the Universe at the cosmological level. It has been well established that the expansion of the Universe is accelerating~\citep{1999ApJ...517..565P,1998AJ....116.1009R}. In an attempt to explain this shortcoming, the $\Lambda$CDM model, which consists of the cold dark matter, cosmological constant, and ordinary matter, was put forward. Despite its success at a large scale, this theory faces severe issues on the sub-galaxy scales, known as the `Small Scale Crisis'~\citep{2017Galax...5...17D}. Moreover, it is flawed with the cosmological constant problem, resulting in disagreements between the observed and theoretically predicted values of the vacuum energy~\citep{1989RvMP...61....1W}. These cosmological observations are thus being increasingly studied using the modified theory of gravity, involving modifications to the Einstein-Hilbert action in GR. 

Proposed by Buchdal, one such class of modified theory of gravity is the $f(R)$ gravity~\citep{2011PhR...509..167C,1970MNRAS.150....1B}, which involves the replacement of $R$ with an arbitrary function $f(R)$ in the Einstein-Hilbert action. In order to derive the modified field equations, one can use either of the two variational principles, each leading to a different formalism. The metric $f(R)$ gravity formalism leads to the fourth-order field equation, on varying the action with respect to the metric. In contrast, the Palatini $f(R)$ gravity leads to a second-order field equation upon varying the action with respect to the metric and the connection independently. The success of any modified theory of gravity lies in the fact that it should be verifiable at all scales and should reduce to GR in the weak-field limit. Therefore, as a testbed, the weak-field limit of $f(R)$ gravity has been rigorously studied in order to compare it with the well-established results of GR at the solar system level~\citep{2012MSAIS..19...63S,2011PhRvD..84b4023E,2021arXiv210501702S}. In recent times, using different forms of $f(R)$ and the model parameter, the $f(R)$ theory is being extensively used to study a wide range of phenomena, starting from inflation~\citep{2018PhRvD..97f4001O,2019PhRvD..99f4049O}, gravitational waves~\citep{2019PhRvD..99l4050K,2021ApJ...909...65K}, to compact objects~\citep{2013JCAP...12..040A,2014PhRvD..89j3509A,2014PhRvD..89f4019G,2009PhRvD..80l4011D}. Moreover, it has been used to address the dark energy problem~\citep{2007IJGMM..04..115N} and also acts as a means of unifying the under- and over-luminous type Ia supernovae (SNe\,Ia) produced from the sub- and super-Chandrasekhar limiting mass white dwarfs (WDs)~\citep{2015JCAP...05..045D,2018JCAP...09..007K,2021IJGMM..1840006W,2017JCAP...10..004B}.

WD marks as the endpoint of stellar evolution for a progenitor star of mass $(10\pm 2) M_\odot$~\cite{2018MNRAS.480.1547L}. They are supported by the electron degeneracy pressure, which acts against the inward force of gravity. However, there is a limit up to which the electron degeneracy pressure can maintain this equilibrium configuration. Combining relativistic energy dispersion with Fermi degeneracy, Chandrasekhar showed that the mass of a non-rotating, non-magnetized WD cannot exceed about $1.4M_\odot$, which is now known as the Chandrasekhar mass-limit~\citep{1935MNRAS..95..207C}. If a WD accumulates matter more than this mass-limit, it undergoes a runaway thermonuclear reaction, leading to an explosion known as the SNe\,Ia~\citep{1997Sci...276.1378N}. Due to this unique mass-limit, SNe\,Ia have nearly similar light curves, and hence, they are used as standard candles. Recently, there have been several observations of peculiar over-luminous~\citep{2006Natur.443..308H,2010ApJ...713.1073S,2009ApJ...707L.118Y} and under-luminous SNe\,Ia~\citep{1992AJ....104.1543F,1998AJ....116.2431T,2001PASP..113..308M,2008MNRAS.385...75T}, which questions the uniqueness of the Chandrasekhar mass-limit. The over-luminous SNe\,Ia were found to have a surprisingly large $^{56}$Ni mass content of upto $1.8M_\odot$, which clearly violates the Khokhlov pure detonation limit~\citep{1993A&A...270..223K}. This suggests that such over-luminous SNe\,Ia cannot originate from WDs following Chandrasekhar mass-limit and thus identifies super-Chandrasekhar limiting mass WDs as their possible candidates. Similarly, the under-luminous SNe\,Ia with $^{56}$Ni mass-content estimate approximately ranging from $0.05M_\odot$ to $0.35M_\odot$ infer indirect evidences of the sub-Chandrasekhar limiting mass WDs~\citep{2006A&A...460..793S}.

Various models were initially propounded for the formation of these peculiar SNe\,Ia. Upon the merger of two sub-Chandrasekhar mass WDs (double-degenerate scenario) to form another sub-Chandrasekhar mass WD and explodes due to the accretion of a helium layer, under-luminous SNe\,Ia can be formed~\citep{2010Natur.463...61P,2000ARA&A..38..191H}. In contrast, several models were proposed to explain the super-Chandrasekhar WDs and thereby the over-luminous SNe\,Ia. Some such models are double-degenerate scenario~\citep{2007ApJ...669L..17H}, presence of magnetic ﬁelds~\citep{2013PhRvL.110g1102D,2014JCAP...06..050D}, presence of a differential rotation~\citep{2012ApJ...744...69H}, presence of charge in the WDs~\citep{2014PhRvD..89j4043L}, ungravity effect~\citep{2016PhRvD..93j4046B}, lepton number violation in magnetized WD~\citep{2015NuPhA.937...17B}, noncommutativity effect~\citep{2021IJMPD..3050034K,2021IJMPD..3050101K}, and many more. Note that these models can at best explain only one class of peculiar SNe\,Ia. Moreover, each of them has some incompleteness, primarily based on stability~\citep{1989MNRAS.237..355K, 2009MNRAS.397..763B}. Even numerical simulations showed that the merger of two massive WDs could never lead to a WD with mass as high as $2.8M_\odot$ due to the off-center ignition and formation of a neutron star rather than an over-luminous SN\,Ia~\citep{2004ApJ...615..444S,2006MNRAS.373..263M}. Hence, most of these pictures failed to explain the full inferred masses ranges for peculiar SNe\,Ia incorporating the same physics.

Das and Mukhopadhyay first initiated the exploration by considering $f(R)=R+\alpha R^2$ gravity with $\alpha$ being the model parameter and showed that this model could explain both the regimes of sub- and super-Chandrasekhar progenitor WDs~\cite{2015JCAP...05..045D}. They showed that negative values of $\alpha$ give super-Chandrasekhar WDs while positive $\alpha$ gives sub-Chandrasekhar limiting mass WDs. Thus one single mass--radius curve could not explain both the mass regimes of WDs. Moreover, $\alpha$ is a fundamental parameter of the model, and varying this parameter to explain similar phenomena is generally not considered to be physical. Furthermore, for this choice of the gravity model, it can be shown that negative $\alpha$ may give rise to a ghost mass~\cite{2016PhRvD..93l4071K,2021ApJ...909...65K}. Hence, Kalita and Mukhopadhyay later considered various higher-order corrections to this model by introducing one more model parameter~\cite{2018JCAP...09..007K}. In this case, the parameters are fixed throughout, and the central density of the WDs, $\rho_\text{c}$ is varied such that low $\rho_\text{c}$ gives sub-Chandrasekhar limiting mass WDs and high $\rho_\text{c}$ gives the super-Chandrasekhar limiting mass WDs. Hence these models can be considered superior to the earlier model. Note that two parameters in $f(R)$ gravity are the minimal requirement to simultaneously explain both the WD mass regimes in the context of general relativistic extensions. However, the lesser the number of parameters, the better the model is. We aim to explore the weak-field regime of $f(R)$ gravity in this context. Because WDs are bigger in size, it is reasonable to approximate them as Newtonian objects. Moreover, besides $f(R)$ gravity, some researchers investigated a few other modified gravity models. Some of them are $f(R,T)$ gravity~\citep{2017EPJC...77..871C}, de Rham-Gabadadze-Tolley (dRGT) like massive gravity~\citep{2019PhRvD..99j4074E}, scalar-vector-tensor gravity (STVG) theory, and Eddington-inspired Born-Infeld gravity~\citep{2017JCAP...10..004B}. However, these models were used to show only one mass-limit of the peculiar WDs (either sub-Chandrasekhar or super-Chandrasekhar). Hence, $f(R)$ gravity can be considered to be a better bet than the other modified gravity theories.

In this paper, we show that only one model parameter survives in the weak-field limit of $f(R)$ gravity, and by fixing this parameter within the suitable bounds, one can obtain both the mass regimes of WDs just by varying $\rho_\text{c}$. This paper is organized as follows. In~\S\ref{Sec: 2}, we discuss the basic equations in $f(R)$ gravity and its weak-field limit. Thereby, we obtain the modified Poisson equation in the Newtonian limit for $f(R)$ gravity. Using this equation, we derive the modified stellar structure equations in~\S\ref{Sec: 3}, which we solve with an appropriate equation of state (EoS) to obtain the mass--radius relation for the WDs in~\S\ref{Sec: 4}. Finally, we put our concluding remarks in~\S\ref{Sec: 5}.

\section{Weak-field limit of $f(R)$ gravity and modified Poisson equation}\label{Sec: 2}
Assuming the sign convention $(-,+,+,+)$, the action for $f(R)$ gravity is given by~\cite{2010LRR....13....3D}
\begin{align}
    S_{f(R)}=\int\left[\frac{c^3}{16 \pi G}f(R)+\mathcal{L}_\mathcal{M}\right]\sqrt{-g}\dd[4]{x},
\end{align}
where $g=\det(g_{\mu\nu})$ is the determinant of the spacetime metric $g_{\mu\nu}$, $c$ is the speed of light, $G$ is Newton's gravitational constant, and $\mathcal{L}_\mathcal{M}$ is the Lagrangian of the matter field. Varying $S_{f(R)}$ with respect to $g_{\mu\nu}$ along with appropriate boundary conditions, the modified field equation in $f(R)$ gravity is given by~\cite{2010LRR....13....3D}
\begin{equation}\label{modified field equation}
    F(R)R_{\mu \nu}-\frac{f(R)}{2}g_{\mu \nu} - \left(\nabla_\mu \nabla_\nu-g_{\mu \nu}\Box \right)F(R) = \kappa T_{\mu \nu},
\end{equation}
where $R_{\mu\nu}$ is the Ricci tensor, $T_{\mu\nu}$ is the energy-momentum tensor, $F(R) = \dv*{f(R)}{R}$, $\kappa = 8\pi G/c^4$, $\Box = -\partial_t^2/c^2 + \laplacian$ is the d'Alembertian operator with $\partial_t$ being the temporal partial derivative and $\laplacian$ the 3-dimensional Laplacian. Note that Greek indices ($\mu$, $\nu$, etc.) take values 0 to 3 where 0 is the temporal component and the rest are the spatial ones, which are denoted by Latin indices ($i$, $j$, $k$, etc.). Now, the trace of Equation~\eqref{modified field equation} is given by
\begin{equation}\label{modified trace equation}
    RF(R)-2f(R)+3\Box F(R) = \kappa g^{\mu \nu} T_{\mu \nu}  = \kappa T.
\end{equation}
In the weak-gravity limit, we assume $g_{\mu\nu}=\eta_{\mu\nu} + h_{\mu\nu}$ and $R = R_0+R_1$, such that $|h_{\mu\nu}|\ll|\eta_{\mu\nu}|$, where $\eta_{\mu\nu}$ is the background Minkowski metric and $R_0$ is the background scalar curvature, while $h_{\mu\nu}$ and $R_1$ are their corresponding tensor and scalar perturbations. Now perturbing Equations~\eqref{modified field equation} and~\eqref{modified trace equation} with these relations, the linearized field and trace equations are given by~\cite{2011PhRvD..83j4022B,2016PhRvD..93l4071K,2021ApJ...909...65K}
\begin{equation}\label{Eq: linearized tensor equation}
    \Box \bar{h}_{\mu\nu} = -\frac{16 \pi G}{c^4}T_{\mu\nu}
\end{equation}
and
\begin{equation}\label{Eq: linearized scalar equation}
    \Box h_f - m^2 h_f = \frac{8 \pi G}{3 F(R_0) c^4} T,
\end{equation}
where 
\begin{equation}\label{Eq: h_bar h relation}
    \bar{h}_{\mu\nu} = h_{\mu\nu} - \left(\frac{h}{2} - h_f \right)\eta_{\mu\nu},
\end{equation}
with $h=\eta_{\mu\nu}h^{\mu\nu}$, $h_f = F'(R_0)R_1/F(R_0)$, and $m$ is the effective mass associated with the scalar degree of freedom, given by 
\begin{equation} \label{Eq: effective mass}
    m^2 = \frac{1}{3}\left[\frac{F(R_0)}{F'(R_0)}-R_0 \right],
\end{equation}
where $F'(R) = \dv*{F}{R}$. Taking the trace of Equation~\eqref{Eq: h_bar h relation}, we obtain
\begin{align}
    \bar{h} = -h + 4h_f \implies h &= -\bar{h}+4h_f.
\end{align}
Now, substituting $\bar{h}=-\bar{h}_{00}+\bar{h}_{ij}$ in the above equation and approximating $\abs{\bar{h}_{00}}\gg \abs{\bar{h}_{ij}}$ for the weak-gravity regime, we obtain
\begin{equation}
    h= \bar{h}_{00}+4h_f.
\end{equation}
Moreover, substituting $\mu=\nu=0$ in Equation~\eqref{Eq: h_bar h relation} together with the above relation, we obtain
\begin{align}\label{Eq: temporal h_bar h relation}
    \bar{h}_{00} = h_{00} + \frac{h}{2}-h_f = h_{00} + \frac{\bar{h}_{00}}{2}+h_f = 2h_{00}+2h_f.
\end{align}
Now, inverting Equation~\eqref{Eq: h_bar h relation} and considering only the spatial components along with the above relation, we obtain
\begin{align}\label{Eq: spatial h_bar h relation}
    h_{ij} &= \bar{h}_{ij} - \left(\frac{\bar{h}}{2}-h_f\right)\eta_{ij} \approx - \left(\frac{\bar{h}}{2}-h_f\right)\eta_{ij} = h_f-\frac{\bar{h}}{2} = h_f + \frac{\bar{h}_{00}}{2} = h_{00}+2h_f.
\end{align}
Therefore, the weak-field metric in $f(R)$ gravity is given by
\begin{align}
    \dd{s^2} = -\left(1-h_{00}\right)c^2\dd{t^2} + \left(1+h_{00}+2h_f\right)\dd{\bm{x}^2}.
\end{align}
In GR, since $h_f=0$ and $h_{00}=-2\phi/c^2$ with $\phi$ being the Newtonian potential, we obtain~\cite{2009igr..book.....R}
\begin{align}\label{Eq: weak field metric}
    \dd{s^2} = -\left(1+\frac{2\phi}{c^2}\right)c^2\dd{t^2} + \left(1-\frac{2\phi}{c^2}\right)\dd{\bm{x}^2}.
\end{align}
This is the well-known weak-field metric in the Newtonian gravity. It is noticed that the due to the influence of $f(R)$ gravity, only the spatial component is modified while temporal one remains unaltered. Now, recognizing $h_{00}=-2\phi/c^2$, the final weak-field metric in $f(R)$ gravity is given by
\begin{align}\label{Eq: modified weak field metric}
    \dd{s^2} = -\left(1+\frac{2\phi}{c^2}\right)c^2\dd{t^2} + \left(1-\frac{2\phi}{c^2}+2h_f\right)\dd{\bm{x}^2}.
\end{align}
Note that the form of $h_f$ depends on the functional form of $f(R)$, which we are going to discuss below.

Now, substituting $\mu=\nu=0$ in Equation~\eqref{Eq: linearized tensor equation} and using the relation of Equation~\eqref{Eq: temporal h_bar h relation}, we obtain
\begin{align}\label{Eq: New tensor equation}
    \Box h_{00} + \Box h_f = -\frac{8\pi G}{c^4} T_{00}.
\end{align}
We now assume perfect non-magnetized fluid in linearized gravity such that $T_{\mu\nu} = \left(\rho + P/c^2\right)u_\mu u_\nu + P \eta_{\mu\nu}$ with $\rho c^2\gg P$, where $\rho$ is the matter density and $P$ is the pressure. Hence, Equations~\eqref{Eq: linearized scalar equation} and~\eqref{Eq: New tensor equation} can be recast as
\begin{align}
    \Box h_f - m^2 h_f &\approx -\frac{8\pi G}{3F(R_0)c^2}\rho\\
    \Box h_{00} + \Box h_f &= -\frac{8\pi G}{c^2} \rho.
\end{align}
Subtracting the first equation from the second one, we obtain
\begin{align}
    \Box h_{00} + m^2 h_f &= -\frac{8\pi     G\rho}{c^2}\left(1-\frac{1}{3F(R_0)} \right).
\end{align}
In this work, we consider the generalized Starobinsky model with $f(R)=R+\alpha R^2+\beta R^3 +\dots$, such that $F(R)=1+2\alpha R+ 3\beta R^2 +\dots$ and $F'(R)=2\alpha+ 6\beta R +\dots$ with $m^2 = 1/6\alpha$ and $h_f = 2\alpha R_1$. Substituting this form of $f(R)$ in the above equation and considering Minkowski background with $R_0=0$, we obtain
\begin{align}\label{Eq: h00 R1 rho relation}
    3\nabla^2 h_{00} + R_1 = -\frac{16\pi G\rho}{c^2}.
\end{align}
Now, for the linearized gravity, we have~\cite{2009igr..book.....R}
\begin{align}
    R_1 = \partial_\mu \partial_\nu h^{\mu\nu} - \Box h.
\end{align}
From Equation~\eqref{Eq: spatial h_bar h relation}, we have $h^{11}=h^{22}=h^{33}=h_{00}+2h_f$ and $h=2h_{00}+6h_f$. Substituting them in the above relation and considering $h_{\mu\nu}$ has no time dependency, we obtain
\begin{align}\label{Eq: R1 h00 relation}
    R_1= -\nabla^2 h_{00} - 8\alpha \nabla^2 R_1.
\end{align}
Substituting $R_1$ in Equation~\eqref{Eq: h00 R1 rho relation} and simplifying, we obtain
\begin{align}
    \nabla^2 h_{00} -4\alpha \nabla^2 R_1 = -\frac{8\pi G\rho}{c^2}.
\end{align}
Again, replacing $R_1$ from Equation~\eqref{Eq: R1 h00 relation}, we obtain
\begin{align}
    \nabla^2 h_{00} +4\alpha \nabla^4 h_{00} +32\alpha^2 \nabla^4 R_1 = -\frac{8\pi G\rho}{c^2}.
\end{align}
It is noticed that Equation~\eqref{Eq: R1 h00 relation} is a recurrence relation. Every time we substitute $R_1$, we obtain higher-order derivatives of $R_1$. Hence, replacing $R_1$ repeatedly, we obtain an infinite series of different derivative orders of $R_1$, which is given by
\begin{align}
    \nabla^2 h_{00} +4\alpha \nabla^4 h_{00} -32\alpha^2 \nabla^6 h_{00} + \dots = -\frac{8\pi G\rho}{c^2}.
\end{align}
Replacing $h_{00}=-2\phi/c^2$ from the aforementioned weak-field metric and simplifying, we obtain
\begin{align}
    \nabla^2 \phi +4\alpha \nabla^4 \phi -32\alpha^2 \nabla^6 \phi + \dots = 4\pi G\rho.
\end{align}
This is field equation in the weak-field regime of $f(R)$ gravity. In the Newtonian case, $\alpha=0$, and we recover the well known Poisson equation, given by $\nabla^2 \phi = 4\pi G\rho$. Hence we call the above equation as the modified Poisson equation in $f(R)$ gravity. It is important to note that only the parameter $\alpha$ appears in the modified Poisson equation, even though $f(R)$ contains more model parameters. This is because $\alpha$ is associated with $R^2$ and it is the only contributing factor for $m$ in Equation~\eqref{Eq: effective mass}.

\section{Modified stellar structure equations}\label{Sec: 3}

In this section, we discuss the stellar structure equations for the compact objects. In Newtonian gravity, because $\nabla^2 \phi = 4\pi G\rho$ holds good, the pressure balance equation is given by~\cite{choudhuri_2010}
\begin{align}
    \dv{P}{r} = -\frac{G M \rho}{r^2},
\end{align}
where $M$ is the mass of the compact object within a radius $r$. Moreover, the mass-estimate equation is given by
\begin{align}\label{Eq: mass estimate}
    \dv{M}{r} = 4\pi r^2 \rho.
\end{align}
Since the Poisson equation is modified, we expect a change in the pressure balance equation. In Equations~\eqref{Eq: h00 R1 rho relation} and~\eqref{Eq: R1 h00 relation}, substituting $h_{00}=-2\phi/c^2$, we obtain
\begin{align}\label{Eq: phi rho R1 equation}
    \nabla^2\phi = \frac{c^2}{6} \left(\frac{16\pi G\rho}{c^2} + R_1\right)
\end{align}
and
\begin{align}
    \nabla^2 \phi = \frac{c^2}{2}\left(8\alpha \nabla^2 R_1 + R_1 \right).
\end{align}
Equating the R.H.S. of these two equations, we obtain
\begin{align}\label{Eq: R1 equation}
    4\alpha \nabla^2 R_1 + \frac{1}{3}R_1 = \frac{8\pi G\rho}{3c^2}.
\end{align}
Moreover, for a fluid parcel moving at a velocity $\bm{v}$ at a time $t$, the Euler equation is given by
\begin{align}
    \pdv{\bm{v}}{t} + \left(\bm{v}\vdot\grad\right)\bm{v} = -\grad{P} -\rho\grad{\phi}.
\end{align}
Now for a spherical symmetric star in hydrostatic equilibrium (i.e. $\bm{v}=0$), this equation reduces to
\begin{align}
\frac{\grad P}{\rho} &= -\grad \phi.    
\end{align}
Taking divergence on both sides and substituting $\nabla^2\phi$ from Equation~\eqref{Eq: phi rho R1 equation}, we obtain
\begin{align}\label{Eq: pressure balance}
\frac{1}{\rho}\nabla^2 P + \grad(\frac{1}{\rho})\vdot\grad P &= -\frac{c^2}{6} \left(\frac{16\pi G\rho}{c^2} + R_1\right).
\end{align}
Equations~\eqref{Eq: mass estimate}, \eqref{Eq: R1 equation}, and~\eqref{Eq: pressure balance} together serve as the hydrostatic balance equations for a compact stars in the weak-field regime of $f(R)$ gravity. Note that the mass-estimate equation is dependent on the form of temporal component of the metric. Comparing Equations~\eqref{Eq: weak field metric} and~\eqref{Eq: modified weak field metric}, it is evident that there is no change in the functional form of $g_{00}$ in the weak-field limit of $f(R)$ gravity, and hence, $\dv*{M}{r}$ equation also remains unaltered. Since we are interested in WDs, we use the Chandrasekhar EoS, given by~\cite{1935MNRAS..95..207C}
\begin{equation}\label{Chandrasekhar EoS}
\begin{aligned}
P &= \frac{\pi m_\text{e}^4 c^5}{3 h^3}\left[x_\text{F}\left(2x_\text{F}^2-3\right)\sqrt{x_\text{F}^2+1}+3\sinh^{-1}x_\text{F}\right],\\
\rho &= \frac{8\pi \mu_\text{e} m_\text{H}(m_\text{e}c)^3}{3h^3}x_\text{F}^3,
\end{aligned}
\end{equation}
where $x_\text{F} = p_\text{F}/m_\text{e}c$, $p_\text{F}$ is the Fermi momentum, $m_\text{e}$ is the mass of electron, $h$ is the Planck's constant, $\mu_\text{e}$ is the mean molecular weight per electron and $m_\text{H}$ is the mass of hydrogen atom. We choose $\mu_\text{e}=2$ indicating the carbon-oxygen WD. Since we consider spherically symmetric WDs, all the variables ($M$, $\rho$, $P$, etc.) depends only on the radial coordinate $r$. Hence in spherical coordinates, using the Chandrasekhar EoS, Equations~\eqref{Eq: mass estimate}, \eqref{Eq: R1 equation}, and~\eqref{Eq: pressure balance} reduce to
\begin{align}\label{Eq: stellar structure 1}
X \dv[2]{\rho}{r} + \frac{2Y}{r} \dv{\rho}{r} - \frac{Y}{\rho} \left(\dv{\rho}{r}\right)^2 &= -\frac{\rho
c^2}{6} \left(\frac{16\pi G\rho}{c^2} + R_1\right), \\ \label{Eq: stellar structure 2}
12\alpha \left(\dv[2]{R_1}{r} + \frac{2}{r} \dv{R_1}{r}\right) + R_1 &= \frac{8\pi G\rho}{c^2}, \\ \label{Eq: stellar structure 3}
\dv{M}{r} &= 4\pi r^2 \rho,
\end{align}
where
\begin{align}
X &= \dv[2]{P}{\rho} = \frac{8 K_2 \left[2+\left(\frac{\rho}{K_1}\right)^{2/3}\right]}{9 K_1^2 \left(\frac{\rho}{K_1}\right)^{1/3}\left[1+\left(\frac{\rho}{K_1}\right)^{2/3}\right]^{3/2}}, \\
Y &= \dv{P}{\rho} = \frac{8 K_2 \left(\frac{\rho}{K_1}\right)^{2/3}}{3 K_1 \sqrt{1+\left(\frac{\rho}{K_1}\right)^{2/3}}},
\end{align}
with
\begin{align}
K_1 = \frac{8\pi \mu_\text{e} m_\text{p} \left(m_\text{e}c\right)^3}{3h^3} \quad {\rm and} \quad K_2 = \frac{\pi m_\text{e}^4c^5}{3h^3}.
\end{align}
Equations~\eqref{Eq: stellar structure 1}-\eqref{Eq: stellar structure 3} needs to be solved simultaneously to obtain the WD structure in the weak-field limit of $f(R)$ gravity.

\section{Mass--radius relation of white dwarfs in the Newtonian regime of $f(R)$ gravity}\label{Sec: 4}

\begin{figure}[htpb]
	\centering
	\includegraphics[scale=0.5]{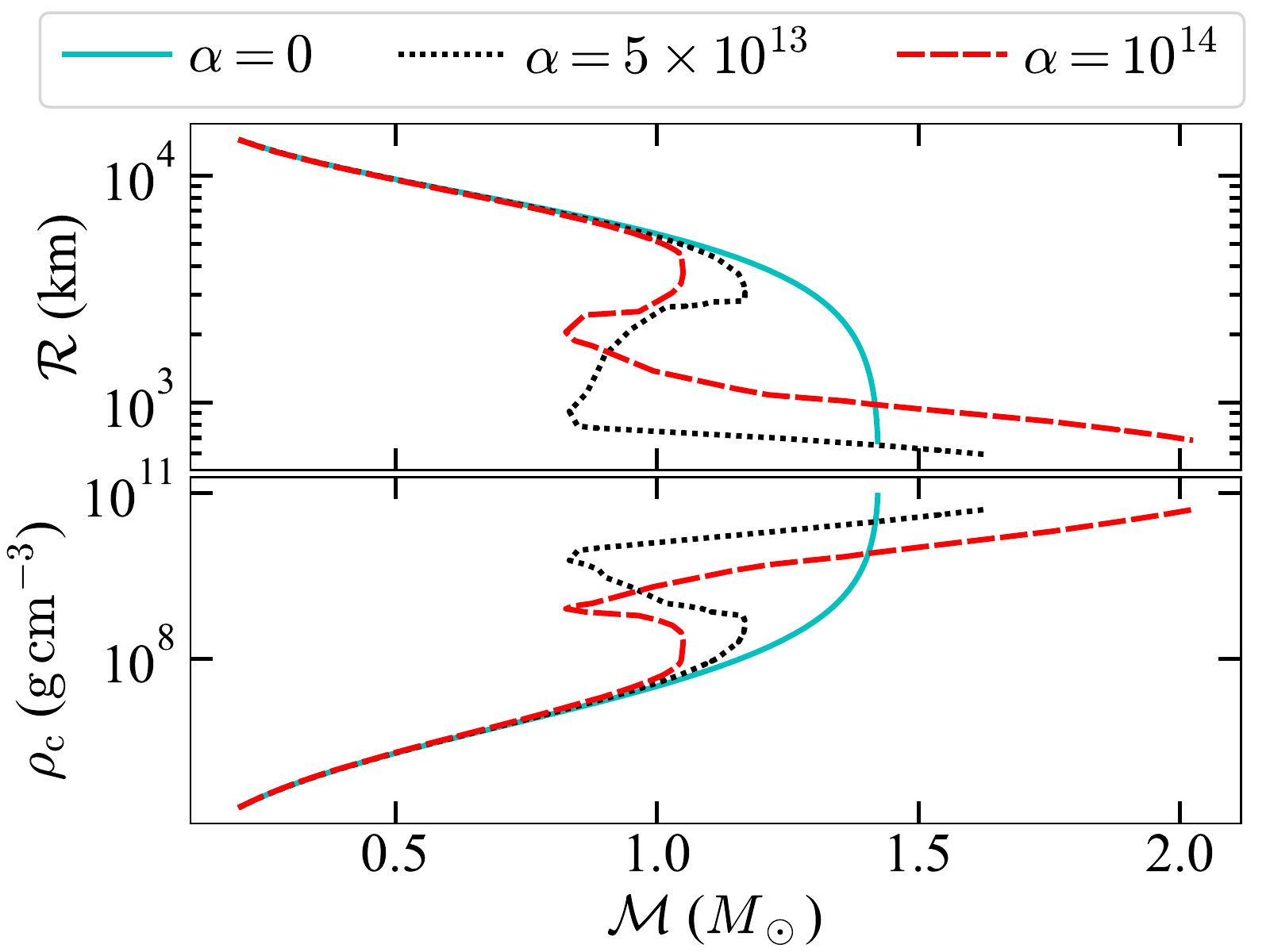}
	\caption{Upper panel: mass--radius relation, Lower panel: variation of WD mass $\mathcal{M}$ with respect to $\rho_\text{c}$ for different values of $\alpha$, which are mentioned as label in cm$^2$ unit.}
	\label{Fig: Mass-radius}
\end{figure}

In this section, we obtain the mass--radius relation of the WDs in the weak-field regime of $f(R)$ gravity. We solve Equations~\eqref{Eq: stellar structure 1}-\eqref{Eq: stellar structure 3} using the fourth-order Runge-Kutta method with the appropriate boundary conditions. We require five boundary conditions since there are two second-order and one first-order differential equations. At the center of the WD, we have
\begin{equation}
    \rho(r=0)=\rho_\mathrm{c}, \quad \left. \dv{\rho}{r}\right \vert_{r=0}=0, \quad M(r=0)=0.
\end{equation}
The two remaining boundary conditions are quite arbitrary. We choose different values of $R_1$ at the center for different values of $\rho_c$, whereas $\dv*{R_1}{r}=0$ at $r=0$. This is because because $R_1$ is nearly proportional to $\rho$ in GR, and since $\dv*{\rho}{r}=0$ at the center, we choose $\dv*{R_1}{r}=0$ there. The values of $R_1$ at the center are chosen in such a way so that the amplitude of $R_1$ decreases from the center to the surface of the WD. The radius of a WD, $\mathcal{R}$ is determined where $\rho$ drops to 0. The mass--radius curve as well as the variation of mass with $\rho_\mathrm{c}$ are shown in Figure~\ref{Fig: Mass-radius} for different values of $\alpha$. 
The bounds on $\alpha$ are chosen from the Gravity Probe B experiment, which states that $\abs{\alpha}\lesssim5\times10^{15}\rm\,cm^2$~\cite{2010PhRvD..81j4003N}. Note that $\alpha=0$ represents the Chandrasekhar mass--radius relation with the limiting mass of about $1.44M_\odot$. For other $\alpha$, the curve overlaps with the Chandrasekhar one at low densities, which means the effect of modified gravity is not significant at such a low density. Also, the values of the model parameter is such that they do not violate the conditions for the solar system test, given by Guo~\cite{2014IJMPD..2350036G}. As the density increases, the curve turns back. It has already been shown that the WDs on the receding branch with $\pdv*{\mathcal{M}}{\rho_\mathrm{c}}<0$ are unstable under radial perturbation, and hence this branch is unstable~\cite{2022PhRvD.105b4028S}. Since unstable branches are nonphysical, the mass corresponding to this peak is the limiting mass of WDs, and it turns out to be sub-Chandrasekhar. Further increase in $\rho_\mathrm{c}$ makes the curve to turn back again, and this branch is now a stable one as $\pdv*{\mathcal{M}}{\rho_\mathrm{c}}>0$. The maximum $\rho_\mathrm{c}$ is chosen in such a way that it is below the neutron drip density and the Chandrasekhar EoS is valid throughout. This stable branch goes beyond the Chandrasekhar mass-limit, and the super-Chandrasekhar WDs are revealed. Thus the modified Poisson equation can explain both the sub- and super-Chandrasekhar limiting mass regimes just by varying $\rho_\mathrm{c}$ of the WDs with once model parameter.

\section{Discussions and conclusions}\label{Sec: 5}

The idea of using modified gravity for sub- and super-Chandrasekhar progenitor WDs to explain respectively the under- and over-luminous SNe\,Ia has been there for quite some time. Among the various modified gravity models, $f(R)$ gravity seems to be one of the prominent bets for this purpose. Initially, the Starobinsky model $f(R)=R+\alpha R^2$ was explored to explain both WD mass regimes~\cite{2015JCAP...05..045D}. However, in this case, both $\alpha$ and $\rho_\mathrm{c}$ need to be varied and hence one single mass--radius curve cannot explain both the peculiar WD mass regimes. It is well-known that negative $\alpha$ gives rise to the imaginary mass~\cite{2016PhRvD..93l4071K}, and hence higher-order corrections to the Starobinsky model, viz. $f(R) = R+\alpha R^2 +\beta R^3 + \dots$ were later proposed~\cite{2018JCAP...09..007K}. In this case, $\alpha$ and $\beta$ were kept fixed within the solar system bound, and just by varying $\rho_\mathrm{c}$, one can obtain the sub- and super-Chandrasekhar limiting mass WDs. Note that in this case, there are two model parameters. In physics, it is generally considered that the lesser the number of model parameters, the better the model is. In this work, we have chosen the weak-field limit of $f(R) = R+\alpha R^2 + \beta R^3 + \dots$. This is because WDs are bigger in size (more than $1000\rm\,km$ in radius), and Newtonian treatment is a good approximation to study their structures. We have found that only the parameter $\alpha$ survives in the weak-field limit, and thereby we have obtained the modified Poisson equation, which was later used to obtain the modified stellar structure equations. We have chosen only positive $\alpha$ to discard any imaginary mass of scalar mode, given by Equation~\eqref{Eq: effective mass}. Note that we do not claim that this model is superior to the previous works in~\citep{2018JCAP...09..007K,2015JCAP...05..045D} just because it has one parameter and a single mass--radius curve explains both the WD mass regime. Rather, unlike the previously mentioned works, which used perturbative approaches, i.e., each physical variable was expanded in terms of their zeroth-order part and one perturbation over them, we have used an unperturbative approach in this work. Thus, in the earlier works, even though those are relativistic calculations, variables like $R$ was expanded as $R = R_0 + \alpha R_1$ and later $R_0$ was replaced by $R_0 = 8\pi G/c^4 \left(\rho_0 c^2 - 3P_0\right)$ in the modified TOV equations. Because the EoS relates $\rho$ and $P$, by supplying density at the center, the value of $R$ is also fixed. On the other hand, in the present work, instead of using such an expression for $R$, we solve a differential equation for $R$ by giving some initial values of $R$. So here $R$ at the center is like a free parameter. It should be kept in mind that the shape of the mass--radius plot depends significantly on the initial values of $R$ in the high-density regime. Hence, in this case, we choose such values of $R(r=0)$ which gives the inverted `S'--shaped curves, to obtain both the sub- and super-Chandrasekhar limiting mass WDs in a single curve. We have shown that in this model, just by varying $\rho_\mathrm{c}$ and keeping $\alpha$ fixed from the Gravity Probe B experiment, one can obtain both the progenitor mass-regime of WDs. Of course, to obtain a more precise mass--radius curve, one needs to solve these equations in the relativistic regime. However, this is not the primary concern for this work, and hence, it is beyond the scope of this paper. In the future, detecting these peculiar WDs through gravitational waves can put better constraints on the various modified gravity theories~\cite{2021ApJ...909...65K}.

\section*{Acknowledgements}
The authors would like to thank the the anonymous reviewer for their comments that have helped to improve the manuscript by providing more insights to the work. SK further thanks Banibrata Mukhopadhyay of IISc for useful discussion of this work.

\bibliographystyle{elsarticle-num}
\bibliography{bibliography}

\end{document}